\pdfoutput=1
\documentclass[twocolumn,showpacs,prb,amsmath,amssymb,floatfix]{revtex4-1}
\usepackage{amsmath,amssymb,color}
\usepackage{bm}
\usepackage{epsfig}
\usepackage{psfrag}

\newcommand{\bd}{\bm}

\begin{document}

\title{Damping of phase fluctuations in superfluid  
Bose gases}

\author{Philipp Lange}
\author{Peter Kopietz}
\affiliation{Institut f\"{u}r Theoretische Physik, Universit\"{a}t
  Frankfurt,  Max-von-Laue Strasse 1, 60438 Frankfurt, Germany}
\author{Andreas Kreisel}
\affiliation{Department of Physics, University of Florida, Gainesville, FL 32611, USA}

\date{\today}
%\date{Received: date / Revised version: date}

 \begin{abstract}
Using Popov's hydrodynamic approach we derive an effective Euclidean 
action  for the  long-wavelength  phase fluctuations of superfluid Bose gases
in $D$ dimensions.
We then use this action to calculate the damping of phase fluctuations
at zero temperature as a function of $D$.
For $D >1$ and wavevectors $| {\bd{k}} | \ll  2 mc$
(where $m $ is the mass of the bosons and $c$ is the sound velocity)
we find that the damping in units of the phonon energy
$E_{\bd{k}} = c | \bd{k} | $ is to leading order
$\gamma_{\bd{k}} / E_{\bd{k}}
= A_D (k_0^D / 2 \pi \rho )  
 ( | \bd{k} |  / k_0 )^{ 2 D -2}$, where
$\rho$ is the boson density and $k_0 =2 mc$ is the inverse healing length. 
For $D \rightarrow 1$ the numerical coefficient $A_D$ vanishes and
the damping is proportional to an additional power of $  |k | /k_0$;
a self-consistent calculation yields in this case $\gamma_{k} / E_k
= 1.32 \,   ( k_0 / 2 \pi \rho  )^{1/2} |k |  / k_0  $.
In one dimension, we also calculate the entire spectral function
of phase fluctuations.
\end{abstract}

\pacs{05.30.Jp, 02.30.Ik, 03.75.Kk}

\maketitle

\section{Introduction}
\label{Intro}

It is well known~\cite{Beliaev58,Gavoret64,Shi98,Andersen04} that the
perturbative treatment of fluctuation corrections
to Bogoliubov's mean-field theory~\cite{Bogoliubov47} for
interacting bosons is plagued by infrared divergencies, which appear at
zero temperature for dimensions $D \leq 3$, and at finite temperature
for $D \leq 4$. 
The physical origin of these divergences is 
the coupling between transverse and longitudinal 
fluctuations~\cite{Nepomnyashchy75,Dupuis11}.
As a consequence, 
the anomalous part of the single-particle 
self-energy $\Sigma_{A} ( 0)$ at vanishing momentum and
frequency is exactly zero \cite{Nepomnyashchy75}, whereas 
Bogoliubov's mean-field theory predicts that $\Sigma_{A} ( 0)$ is finite.
To recover the exact result $\Sigma_{A} ( 0)=0$ diagrammatically,
infinite orders have to be re-summed using non-perturbative methods, such as
the renormalization group.~\cite{Castellani97,Sinner09,Dupuis09}

If one is interested in long-wavelength and low-energy
properties of the system, Popov's quantum hydrodynamic 
approach\cite{Popov72,Popov83}
offers an alternative parametrization of the fluctuations which does not lead to
infrared divergencies. In this approach one separates the low-energy from the high-energy
modes and treats the  low-energy sector within a gradient expansion for the
phase and amplitude fluctuations.
This hydrodynamic approach can also be used to study
interacting bosons in one spatial dimension,
where strong fluctuations prohibit the formation of a 
Bose-Einstein condensate\cite{Popov83,Popov77}, although
the groundstate is superfluid.
In fact, in one dimension the weak coupling expansion of thermodynamic
quantities obtained within the hydrodynamic approach 
agrees with exact results for the Lieb-Liniger model\cite{Lieb63}
up to the second order in the
relevant dimensionless interaction parameter\cite{Popov77}.
On the other hand, the single-particle spectral function and
the dynamic structure factor (spectral function for density fluctuations)
of interacting bosons in one dimension have recently been shown
to exhibit algebraic singularities.\cite{Khodas07,Imambekov08} 
In principle it should be possible to reproduce these singularities within
the hydrodynamic approach, but this requires a non-perturbative
treatment of the interactions between amplitude and phase fluctuations
which is beyond the scope of this work.

Here we shall use
the hydrodynamic approach to calculate
the damping of phase fluctuations
in low dimensional Bose gases.
In one dimension we also calculate the
entire spectral function of phase fluctuations and show that
in the vicinity of the phonon peaks it 
has approximately Lorentzian line-shape,
with on-shell damping proportional to $k^2$ for small wavevectors $k$.
We also elaborate on the relation between the $k^2$-scaling of the
damping in $D=1$ and the
Beliaev damping of the phonon mode in superfluid Bose gases, which
in $D >1$ is known to scale as
$| \bd{k} |^{2 D -1}$ for small wavevectors \cite{Kreisel08}.

\section{Effective action for phase fluctuations}
\label{Effective_action}
According to Popov\cite{Popov83} the long-wavelength asymptotics of
correlation functions of interacting bosons can be obtained from
an effective long-wavelength
hydrodynamic action 
involving a phase field $\varphi ( \bd{r} , \tau )$ and a conjugate density field
$\rho ( \bd{r} , \tau )$.
These are slowly varying functions
of space $\bd{r}$ and the imaginary time $\tau$, and
are defined by writing the slowly varying
part  of the fundamental boson field as
 \begin{equation}
 \psi ( \bd{r} , \tau )  = \sqrt{ \rho ( \bd{r} , \tau )} e^{ i \varphi ( \bd{r} , \tau ) }.
 \end{equation}
Setting $\rho ( \bd{r}, \tau ) = \rho_0 + \sigma ( \bd{r} , \tau )$, where
 \begin{equation}
 \rho_0 = \frac{1}{\beta V}\int d^D r \int d \tau \rho ( \bd{r} , \tau )
 \end{equation}
is the spatial and temporal average of the density field, and expanding
the effective action of the slow part of the boson field
to second order in the gradients, we obtain the
hydrodynamic Euclidean action for the slowly varying phase and amplitude
fluctuations~\cite{Popov83}
 \begin{eqnarray}
 S  [ \varphi, \sigma  ] = - \beta V p   ( \mu ,  \rho_0 )
 + S_{2} [ \varphi, \sigma ],
 \end{eqnarray}
where $\beta$ is the inverse temperature, $V$ is the volume of the system,
and $p ( \mu , \rho_0 )$ is the pressure
as a function of the chemical potential $\mu$ and 
the average density $\rho_0$, and
$S_{2} [ \varphi, \sigma ]$ contains fluctuation corrections
up to second order in the derivatives,
 \begin{eqnarray}
 S_{2} [ \varphi, \sigma ] 
 & = & \int_0^{\beta} d \tau \int d^D r
 \Bigl[  p_{\mu}   \frac{ (\nabla \varphi)^2}{2m} +
 p_{\mu \mu } \frac{ ( \partial_{\tau } \varphi)^2}{2} 
 \nonumber
 \\
&  & \hspace{-10mm}  - i p_{\mu \rho_0}  \sigma \, \partial_{\tau} \varphi  -
  p_{\rho_0 \rho_0} \frac{ \sigma^2}{2} 
%\nonumber
% \\
%& &
+  \frac{ (\nabla \sigma)^2}{8m \rho_0}
+ \frac{  ( \nabla \varphi )^2   \sigma }{2m}
   \Bigr].
 \nonumber
 \\
 & &
 \label{eq:Seff}
 \end{eqnarray}
Here $m$ is the mass of the bosons and
the coefficients $ p_{\mu}$, $p_{\mu  \mu }$,
$p_{\rho_0 \rho_0}$ and $p_{\mu \rho_0}$ are the
partial derivatives of the pressure $p ( \mu , \rho_0 )$
of a homogeneous system with chemical potential $\mu$ and 
average density $\rho_0$. The
last two terms on the right-hand side of Eq.~(\ref{eq:Seff}) represent the kinetic energy
of the slowly oscillating part of the boson field.
A simple approximation for the pressure is \cite{Popov83} 
 \begin{eqnarray}
 p ( \mu , \rho_0 ) & \approx & \mu \rho_0 - \frac{u_0}{2} \rho_0^2
% \nonumber
% \\
% & = &
 = - \frac{u_0}{2} \left[ ( \rho_0 - {\rho} )^2 - {\rho}^2 
 \right], \hspace{7mm}
 \label{eq:pmu}
 \end{eqnarray}
where $u_0$ is the two-body interactions in vacuum for vanishing external momenta \cite{Olshanii98}, and
${\rho} = \mu / u_0$ is the value of
the fluctuating variable $\rho_0$ at the saddle point of the functional 
integral. In the thermodynamic limit and at zero temperature both $\rho$ and $\rho_0$ can be identified with the total
density of the bosons.  
Eq.~(\ref{eq:pmu}) implies
the following estimate for the relevant partial derivatives of the pressure,
 \begin{subequations}
 \begin{eqnarray}
 p_{\mu} & \approx & \rho = \mu / u_0,
 \label{eq:p1}
 \\
 p_{\mu \mu } & \approx & 0,
 \label{eq:p2}
 \\
 p_{\mu \rho_0} & \approx & 1,
 \label{eq:p3}
\\
 p_{\rho_0 \rho_0} & \approx & - u_0.
\label{eq:p4}
 \end{eqnarray}
\end{subequations}
The above hydrodynamic action describes long-wavelength fluctuations at length scales
larger than some cutoff scale $1/ \Lambda_0$.
In momentum space we should therefore impose an ultraviolet cutoff $\Lambda_0$
on all integrations. In the weak coupling regime
a reasonable choice of the cutoff is the inverse healing length
$\Lambda_0 = 2 m c$, where $c$ is the sound velocity defined below.

Introducing the Fourier transform of the 
fields in  momentum-frequency space,
 \begin{subequations} 
 \begin{eqnarray}
 \varphi ( {\bd{r}} , \tau ) & = & \int_K e^{ i (\bd{k} \cdot {\bd{r}} -  \omega \tau)} \varphi_K,
 \\ 
 \sigma ( {\bd{r}} , \tau ) &  =  & \int_K e^{ i (\bd{k} \cdot {\bd{r}} -  \omega \tau)} \sigma_K,
\end{eqnarray}
 \end{subequations}
that the gradient contribution (\ref{eq:Seff})  to the hydrodynamic action
can be written as
 \begin{eqnarray}
 S_{2}  [ \varphi, \sigma ] & = & \frac{1}{2}\int_K \Biggl[ 
\Bigl(   \frac{p_{\mu}}{m} \bd{k}^2   +p_{\mu \mu}    \omega^2  \Bigr) 
\varphi_{-K}  \varphi_K  
 \nonumber
 \\
 & & \hspace{8mm}  + p_{\mu \rho_0} \omega\,  ( \varphi_{-K} \sigma_K - \sigma_{-K} 
\varphi_{K} )
 \nonumber
 \\
 & & \hspace{8mm} + \Bigl( - p_{\rho_0 \rho_0} + \frac{\bd{k}^2}{4 m \rho_0}  \Bigr) 
\sigma_{-K}  \sigma_K  
\Biggr]
 \nonumber
 \\
 &  & \hspace{-16mm} - \frac{1}{2} \int_{ K_1} \int_{K_2} \int_{K_3} \delta_{K_1 + K_2 + K_3,0}
 \frac{ \bd{k}_1 \cdot \bd{k}_2}{m} \varphi_{ K_1} \varphi_{K_2} \sigma_{K_3}. 
 \nonumber
 \\
 & &
 \label{eq:SeffFourier}
 \end{eqnarray}
Here $K = ( \bd{k} , i \omega )$ is a collective label for
momenta $\bd{k}$ and bosonic Matsubara frequencies $ i \omega$,
the integration symbols represent
$\int_K = ( \beta V )^{-1} \sum_{\bd{k}} \sum_{\omega }$, and
the normalization of the delta-symbols is
$\delta_{ K , K^{\prime}} = \beta V \delta_{\bd{k} ,
 \bd{k}^{\prime} } \delta_{ \omega , \omega^{\prime}}$
where the $\delta$-symbols on the right-hand side are  Kronecker-deltas.
Since the hydrodynamic action (\ref{eq:SeffFourier}) is quadratic in the 
amplitude field $\sigma$, we may carry out the functional integration
over the $\sigma$-field,
 \begin{equation}
 e^{ - S_{\rm eff} [ \varphi ] } = \int {\cal{D}} [ \sigma ]
   e^{ - S_2 [ \varphi, \sigma ] }.
 \end{equation}
The effective action of the phase field is
 \begin{eqnarray}
 S_{\rm eff} [ \varphi ] & = &  \frac{1}{2} \int_K G_0^{-1} ( K ) 
 \varphi_{-K } \varphi_K
 \nonumber
 \\
 & +& \frac{1}{3!} \int_{ K_1} \int_{ K_2} \int_{K_3} \delta_{ K_1 + K_2 + K_3 , 0}
 \nonumber
 \\
 & & \times  \Gamma^{(3)}_0 ( K_1 , K_2 , K_3 ) 
\varphi_{K_1} \varphi_{K_2} \varphi_{K_3}
 \nonumber
 \\
 & +& \frac{1}{4!} \int_{ K_1} \int_{ K_2} \int_{K_3} \int_{K_4} 
\delta_{ K_1 + K_2 + K_3 + K_4, 0}
 \nonumber
 \\
 & & \times  \Gamma^{(4)}_0 ( K_1 , K_2 , K_3 ,K_4) 
\varphi_{K_1} \varphi_{K_2} \varphi_{K_3} \varphi_{K_4},
 \hspace{7mm}
 \label{eq:Seffphase}
 \end{eqnarray} 
where the inverse Gaussian propagator of the phase field is
 \begin{equation}
 G_0^{-1} (K ) = \frac{ p_{\mu}}{m} \bd{k}^2 +
 \left[ p_{\mu \mu} + \frac{ {p}^2_{\mu \rho_0}}{- p_{\rho_0 \rho_0} +
 \frac{ \bd{k}^2}{ 4 m \rho_0 }} \right] \omega^2,
 \end{equation}
and the properly symmetrized three-point and four-point vertices are
 \begin{eqnarray}
  \Gamma^{(3)}_0 ( K_1 , K_2 , K_3 ) & = & - p_{\mu \rho_0}
 \frac{ \bd{k}_1 \cdot \bd{k}_2}{m} \frac{ \omega_3}{ - p_{\rho_0 \rho_0} + 
 \frac{ \bd{k}_3^2}{4 m \rho_0} } 
 \nonumber
 \\
 & & \hspace{-7mm} + 
 ( K_2 \leftrightarrow K_3) +
 ( K_1 \leftrightarrow K_3 ),
 \label{eq:Gamma3def}
\\
 \Gamma^{(4)}_0 ( K_1 , K_2 , K_3, K_4 ) & = & -
 \frac{ ( \bd{k}_1 \cdot \bd{k}_2) (  \bd{k}_3 \cdot \bd{k}_4)     }{m^2 \left(
 - p_{\rho_0 \rho_0} + 
 \frac{ ( \bd{k}_1+ \bd{k}_2)^2}{4 m \rho_0} \right) } 
 \nonumber
 \\
 & &\hspace{-7mm}  +
 ( K_2 \leftrightarrow K_3) +
 ( K_2 \leftrightarrow K_4 ).
 \label{eq:Gamma4def}
 \end{eqnarray}
Note that the non-Gaussian contributions to the
effective hydrodynamic action (\ref{eq:Seffphase})
of the phase fluctuations are generated by the
term $( \nabla \varphi )^2  \sigma / (2 m )$
associated with the coupling between amplitude and phase fluctuations
in our original hydrodynamic action (\ref{eq:Seff}).

\section{Damping of phase fluctuations in  dimensions $D > 1$}
\label{Damping}

Within the Gaussian approximation we obtain the 
energy dispersion $E_{\bd{k}}$ of the phase fluctuations from the
condition $G_0^{-1} ( \bd{k} , E_{\bd{k}} + i \eta ) =0$.
Approximating the pressure derivatives by Eqs.~(\ref{eq:p1}--\ref{eq:p4})
we obtain 
 \begin{equation}
 G_0 ( K ) = \frac{u_0 ( 1 + \bd{k}^2 / k_0^2 )}{ \omega^2 + E_{\bd{k}}^2 },
 \label{eq:G0def}
 \end{equation}
where $E_{\bd{k}}$ is
the Bogoliubov dispersion,
 \begin{equation}
 E_{\bd{k}} = c | \bd{k} | \sqrt{ 1 + \bd{k}^2 / k_0^2 }.
 \end{equation}
Here the sound velocity is given by
 \begin{equation}
 c = \sqrt{ \frac{u_0 \rho}{m}} = \sqrt{ \frac{\mu}{m}},
 \end{equation}
and the inverse healing length
 \begin{equation}
 k_0 = 2 m c = 2 \sqrt{  m \mu }
 \label{eq:k0def}
 \end{equation}
marks the crossover from the linear regime
of a sound-like dispersion
to the quadratic regime of quasi-free bosons.
Note that the bare coupling can be written as
 \begin{equation}
 u_0 = \frac{ mc^2}{\rho} ,
 \end{equation}
which in one dimension has units of velocity.
In fact, in $D=1$ the dimensionless ratio $u_0/c = mc / \rho$
can be identified with the usual Lieb-Liniger parameter \cite{Lieb63} which
is the  relevant dimensionless interaction strength.

\begin{figure}[tb]
  \centering
 \includegraphics[width=0.3\textwidth]{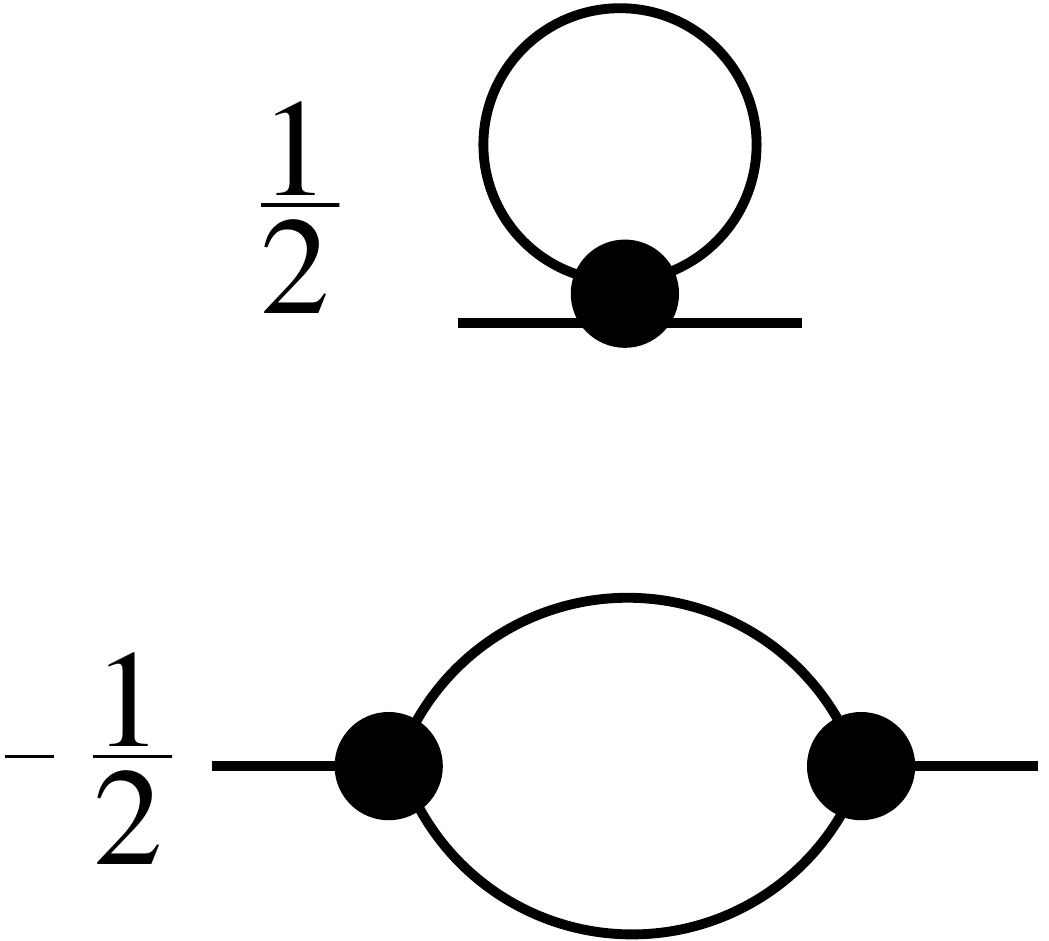}
  \caption{
These Feynman diagrams represent the 
first two perturbative corrections to the self-energy
of the phase fluctuations, see Eqs.~(\ref{eq:Sigma1}, \ref{eq:Sigma2}).
The solid lines represent the Gaussian propagator
$G_0 ( K )$ given in Eq.~(\ref{eq:G0def}), while
the black dots denote the symmetrized three-point and four-
point vertices defined in Eqs.~(\ref{eq:Gamma3def}, \ref{eq:Gamma4def}).
}
\label{fig:Feynman}
\end{figure}

\begin{figure}[tb]
  \centering
\includegraphics[width=0.45\textwidth]{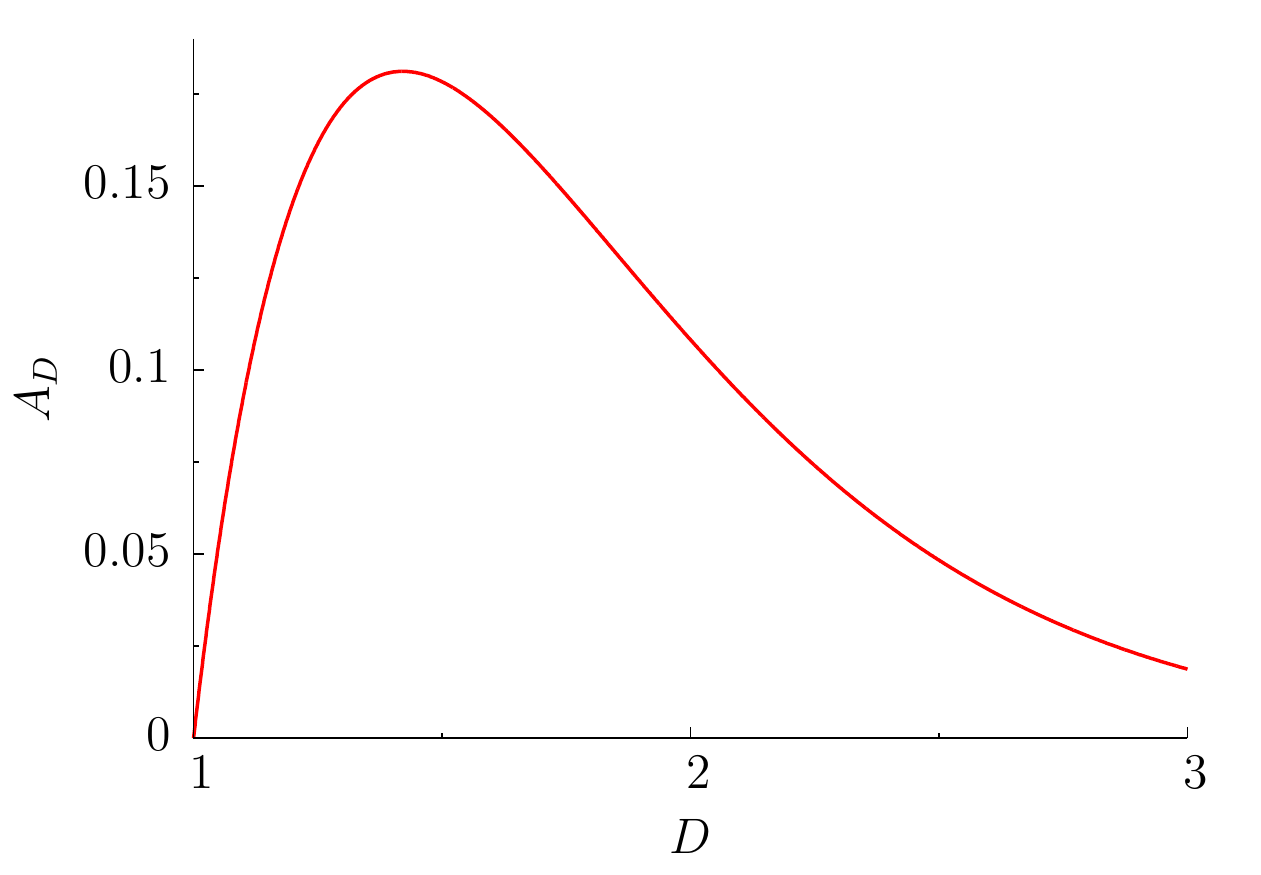}%\nspace
  \caption{%
%(Color online)
Graph of the numerical coefficient $A_D$ defined in
Eq.~(\ref{eq:alphaDdef}) as a function of $D$ for  $1 \leq D \leq 3$.
}
\label{fig:alphaD}
\end{figure}

The interactions in our effective action (\ref{eq:Seffphase})
give rise to a momentum- and frequency dependent self-energy
$\Sigma ( K )$, so that the true
inverse propagator of the phase fluctuations is
 \begin{equation}
 G^{-1} ( K ) =  G_0^{-1} ( K ) + \Sigma (K ).
 \end{equation}
The renormalized energy dispersion $\tilde{E}_{\bd{k}}$
of the phase mode and its damping $\gamma_{\bd{k}}$ are
given by the solutions of $G^{-1} ( \bd{k} , \tilde{E}_{\bd{k}} + i  \gamma_{\bd{k}} ) =0$.
To first order in the quartic vertex $\Gamma_0^{(4)}$ and to second order
in the cubic vertex $\Gamma_0^{(3)}$ the self-energy is
$ \Sigma ( K ) = \Sigma_1 ( K ) + \Sigma_2 ( K )$, where
 \begin{eqnarray}
 \Sigma_1 ( K ) & = & \frac{1}{2} \int_{ K^{\prime}} G_0 ( K^\prime )
 \Gamma_0^{(4)} ( K^{\prime} , - K^{\prime} , K  , -K ),
 \label{eq:Sigma1}
 \\
 \Sigma_2 ( K )  & =   & - \frac{1}{2} \int_{ K^{\prime}} G_0 ( K^{\prime} ) G_0 ( K^{\prime} + K )
 \nonumber
 \\
 & & \hspace{-15mm} \times \Gamma_0^{(3)} ( K , - K - K^{\prime} , K^{\prime} )
  \Gamma_0^{(3)} (  - K^{\prime} , K +  K^{\prime} ,  -K  ).
 \hspace{7mm}
 \label{eq:Sigma2}
 \end{eqnarray}
The corresponding Feynman diagrams are shown
in Fig.~\ref{fig:Feynman}.
To lowest order in perturbation theory the
damping of the phase mode is given by
 \begin{equation}
 \gamma_{\bd{k}} = - \frac{u_0 ( 1 + {\bd{k}}^2 / k_0^2 )}{2 E_{\bd{k}}}
 {\rm Im} \Sigma_2 ( \bd{k} , E_{\bd{k}} + i \eta ),
 \label{eq:ImS}
 \end{equation}
where $ \eta >0$ is infinitesimal.
Substituting $\Sigma_2 ( K)$ from Eq.~(\ref{eq:Sigma2})
into Eq.~(\ref{eq:ImS}) and using
Eqs.~(\ref{eq:G0def}) and (\ref{eq:Gamma3def}) for
$G_0 $ and $\Gamma_0^{(3)}$,
we obtain for $ |\bd{k} | \ll k_0$ after straightforward algebra
 \begin{equation}
 \gamma_{\bd{k}} = \frac{\pi u_0}{16 m^2} \int \frac{d^{D} k^{\prime}}{ (2 \pi )^D}
 \delta (  E_{\bd{k}}  -   E_{\bd{k}^{\prime}} -    E_{ \bd{k} - \bd{k}^{\prime}} ) 
  W_{ \bd{k} ,  \bd{k}^{\prime} },
 \end{equation}
with
 \begin{eqnarray}
  W_{ \bd{k} ,  \bd{k}^{\prime} } & = & 
 \frac{ [ \bd{k}^{ 2} - \bd{k}^{\prime 2} ]^2}{E_{\bd{k} -
 \bd{k}^{\prime}}} +
 \frac{ [ \bd{k}^{ 2} - ( \bd{k} - \bd{k}^{\prime })^2 ]^2}{E_{\bd{k}^{\prime}}}
 \nonumber
 \\
  & & - \frac{ [ \bd{k}^{\prime 2} - ( \bd{k} - \bd{k}^{\prime })^2 ]^2}{E_{\bd{k}}}.
 \end{eqnarray}
Taking into account that the function 
  $W_{ \bd{k} ,  \bd{k}^{\prime} }$ is multiplied by
$ \delta (  E_{\bd{k}}  -   E_{\bd{k}^{\prime}} -    E_{ \bd{k} - \bd{k}^{\prime}} ) $,
we may substitute under the integral sign for small momenta
 \begin{equation}
 W_{ \bd{k} ,  \bd{k}^{\prime} } \rightarrow \frac{9}{c} | \bd{k} | | \bd{k}^{\prime} |
 | \bd{k} - \bd{k}^{\prime} |.
 \end{equation}
The $\bd{k}^{\prime}$-integration can now be performed using $D$-dimensional
spherical coordinates. For small external momentum
$\bd{k}$ the loop momentum $\bd{k}^{\prime}$ is almost parallel to
$\bd{k}$ so that we may approximate \cite{Kreisel08}
 \begin{equation}
  \delta (  E_{\bd{k}}  -   E_{\bd{k}^{\prime}} -    E_{ \bd{k} - \bd{k}^{\prime}} )
 \approx \frac{k_0 \delta ( \vartheta - 
 \sqrt{3} \frac{ | \bd{k} |  - | \bd{k}^{\prime} |}{ k_0}  )  }{
 \sqrt{3} c | \bd{k}|  | \bd{k}^{\prime} | },
 \end{equation}
where $\vartheta$ is the angle  between $\bd{k}$ and $\bd{k}^{\prime}$.
We finally obtain in $D$ dimensions
 \begin{equation}
 \frac{ \gamma_{\bd{k}}}{ E_{\bd{k}} } = 
 A_D \frac{ k_0^D}{2 \pi \rho} 
 \left( \frac{ | \bd{k} | }{ k_0 } \right)^{ 2( D-1) },
 \label{eq:Beliaev}
 \end{equation}
where the numerical coefficient $A_D$ can be expressed in terms of
$\Gamma$-functions as follows,
 \begin{equation}
 A_D = \frac{ 3^{ \frac{D+1}{2}}  \pi^{ 2 - \frac{D}{2}  }    \Gamma ( D )    }{ 2^{  3D}
  \Gamma ( \frac{ D-1}{2} ) \Gamma ( D + \frac{1}{2} )}.
 \label{eq:alphaDdef}
 \end{equation}
A graph of  $A_D$ as a function of the dimensionality  $D$ of the system
is shown in Fig.~\ref{fig:alphaD}.
In $D=3$ and $D=2$ we obtain
 $A_3 = \frac{3}{160} \approx 0.0187$ and 
 $A_2 = \frac{\sqrt{3} }{16} \approx 0.108 $, while
$A_D \sim \frac{3 \pi}{8} ( D-1 ) \rightarrow 0$ for $D \rightarrow 1$.
In three dimensions Eq.~(\ref{eq:Beliaev}) agrees with the 
well-known Beliaev damping of 
the phonon mode in a Bose condensate \cite{Beliaev58}. 
Beliaev damping in $D=3$ and $D=2$ has recently been re-derived
in Ref. \cite{Chung09} using Popov's hydrodynamic approach;
however, these authors did not integrate out the amplitude fluctuations,
which renders the algebra more complicated than in our approach based on the
effective action of phase fluctuations.
The fact that for arbitrary $D > 1$ 
Beliaev  damping scales as $| \bd{k}|^{2 D-1}$ has been pointed out
previously by several authors \cite{Kreisel08,Sinner09,Zhitomirsky12}.

\section{Phase fluctuations in one dimension}
\label{Phase_fluc_1D}
Obviously, Eq.~(\ref{eq:Beliaev}) cannot be used to estimate 
the damping of phase fluctuations in one dimension, because
the coefficient $A_D$ vanishes for $D \rightarrow 1$.
The problem is that in the derivation of  Eq.~(\ref{eq:Beliaev})
we have inserted  bare Green functions in the loop integration, which 
in $D=1$ is
not accurate enough 
to obtain the damping of the phase fluctuations. 
A similar problem arises in the calculation
of the damping of the excitations of a clean Luttinger liquid, which has been
studied  by Samokhin \cite{Samokhin98}
by means of a self-consistent perturbative calculation taking the damping
of intermediate states into account.
Although in this case the spectral function is known to exhibit a non-Lorentzian
line-shape with algebraic singularities \cite{Pustilnik06}, 
the overall
width of the spectral function can be estimated correctly with this method.

\begin{figure}[tb]
  \centering
\includegraphics[width=0.45\textwidth]{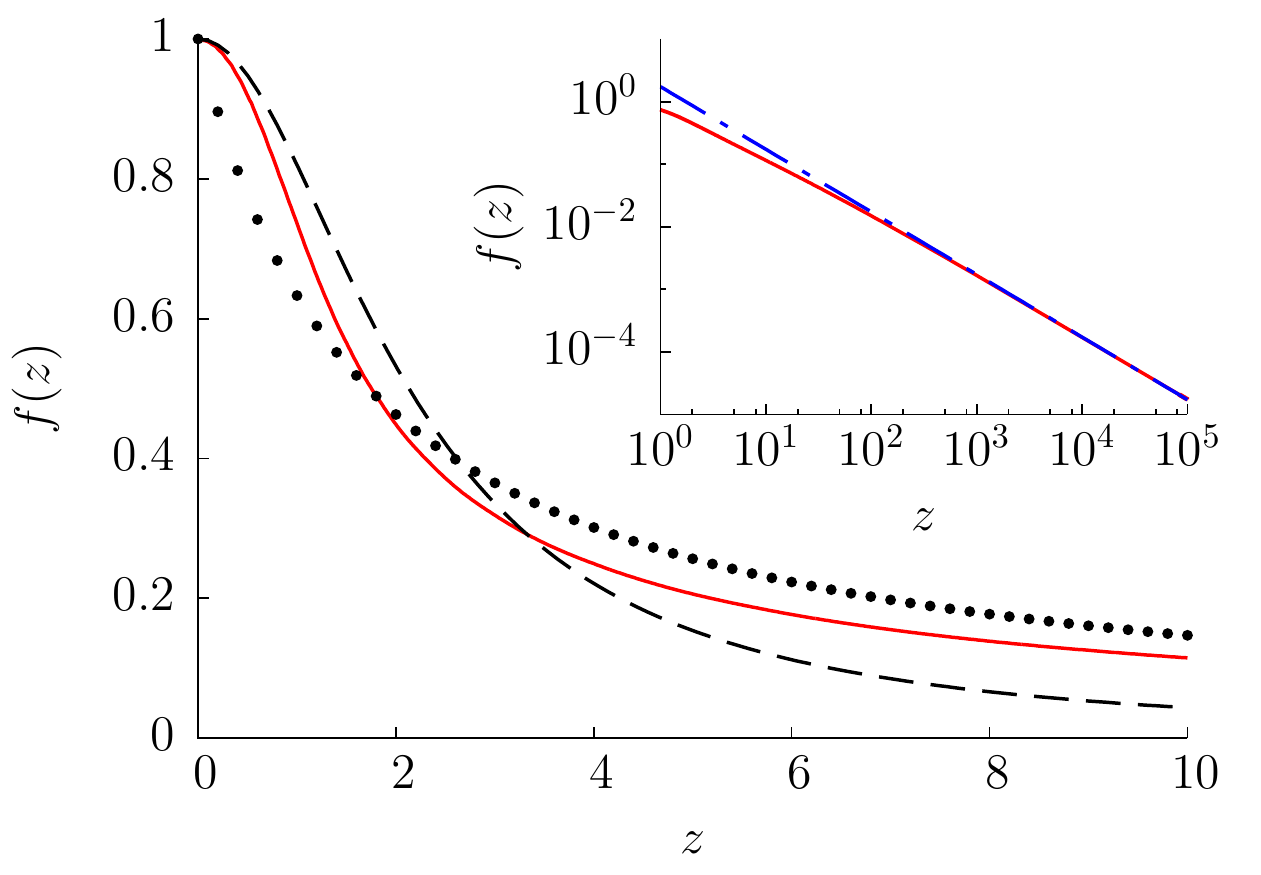}
  \caption{
The solid line is the numerical solution of the
integral equation (\ref{eq:integral}).
The dashed line represents the approximation (\ref{eq:ansatz}) with $f_2 = 0.22$,
which is reasonably accurate for $ | z | \lesssim 5$.
The dotted line represents the interpolation (\ref{eq:fasym}) with
$f_1 = 0.58$.
The logarithmic plot in the inset shows that  for large $|z|$ the
solution of the integral equation (\ref{eq:integral})  vanishes as $1 /  | z |  $.
The dashed-dotted line in the inset is the curve $1/( 0.58  | z |) $.
}
\label{fig:fint}
\end{figure}

Let us now use the method proposed by Samokhin \cite{Samokhin98} (see also Ref. \cite{Andreev80})
to calculate the damping of phase fluctuations in the one-dimensional
Bose gas. In fact, we shall go beyond Samokhin's work and calculate
the entire spectral line-shape of phase fluctuations. 
To include the damping of intermediate states
in our perturbative self-energy (\ref{eq:Sigma2}), we 
replace the Gaussian propagators on the
right-hand side by the exact propagators $G ( K )$ of the phase mode,
for which we use the  spectral representation
 \begin{equation}
  G ( k , i \omega ) = \int_{- \infty}^{\infty}
 \frac{ d \omega^{\prime}}{2 \pi} \frac{B ( k , \omega^{\prime} )}{
  \omega^{\prime 2} + \omega^2 },
 \end{equation}
where the spectral function
 \begin{equation}
 B ( k  , \omega ) = 2 \omega {\rm Im} G ( k , \omega + i \eta )
 \end{equation}
is real and positive.
Retaining only the imaginary part of the self-energy, we find
after analytic continuation that
for frequencies close to $\pm E_{k}$ the spectral function
can be approximated by
 \begin{equation}
 B ( k , \omega  ) \approx 
 \frac{u_0  \gamma ( k , \omega )}{( | \omega | - E_k )^2 + \gamma^2 (
 k, \omega )} .
 \end{equation}
where the damping function $\gamma ( k , \omega)$ satisfies 
the integral equation
 \begin{eqnarray}
 \gamma ( k , \omega ) & = & \frac{{\rm sgn} \omega   }{16 m^2 u_0}
  \int \frac{d k^{\prime}}{2 \pi}
  \int_0^{ | \omega |} d \omega^{\prime} B ( k^{\prime} , \omega^{\prime} )
 \nonumber
 \\
 & & \hspace{-10mm} \times 
 B ( k - k^{\prime} , \omega - \omega^{\prime} )
 \biggr\{
 \frac{ [ {k}^{ 2} - {k}^{\prime 2} ]^2}{ | \omega |
   - \omega^{\prime}}
 \nonumber
 \\
  & & \hspace{-10mm}
 +
 \frac{ [ {k}^{ 2} - ( {k} - {k}^{\prime })^2 ]^2}{
   \omega^{\prime}}
 - \frac{ [ {k}^{\prime 2} - ( {k} - {k}^{\prime })^2 ]^2}{| \omega |}  \biggl\}  .
 \label{eq:selfcon}
 \end{eqnarray}
To solve this
non-linear integral equation, we make the ansatz~\cite{Samokhin98}
 \begin{equation}
 \gamma ( k , \omega ) = \gamma_k f \left( \frac{| \omega | - E_k}{\gamma_k }
 \right),
 \label{eq:gammascaling}
 \end{equation}
where the on-shell damping is assumed to be of the form
 $\gamma_k = f_0 | k |^{\alpha}$, with some exponent $\alpha$. 
The dimensionless function
$f (z )$  is normalized such that $f(0)=1$, so that
the dimensionful constant $f_0$ determines the 
strength of the on-shell damping.
The function $f ( z )$ is
expected to be strongly peaked to $z=0$  and to decay as a power law for
$| z | \gg 1$.
After substituting the ansatz (\ref{eq:gammascaling}) into  
Eq.~(\ref{eq:selfcon})  we may scale out the $k$-dependence by
introducing  dimensionless integration variables
$x = k^{\prime} / k$ and
$y =  ( \omega^{\prime} - E_{k^{\prime}}) / \gamma_{ k^{\prime}}$.
It is then easy to see that
our ansatz is only consistent if $\alpha =2$. The constant $f_0$ is then given by
 \begin{equation}
 f_0 = \frac{3  \sqrt{ I_0 [ f ]}  }{4m}  \sqrt{ \frac{k_0}{ 2 \pi \rho} } ,
 \end{equation}
where the function $f (z )$ satisfies the integral equation
\begin{equation}
    f (z ) = \frac{ I_z [f]}{  I_0 [ f ] }    ,
 \label{eq:integral}
 \end{equation}
with the non-linear functional $I_z [ f ]$ given by
 \begin{eqnarray}
 I_z [ f ] & = &  \int_{0}^1 dx \frac{x}{1-x}
 \int_{ - \infty}^{\infty} dy \frac{ f(y)}{y^2 + f^2 ( y )}
 \nonumber
 \\
 & & \times   
\frac{ f \left(  \frac{(z-y)x^2}{(1-x)^2}   \right)}{ 
\bigl[ \frac{(z-y) x^2 }{(1-x)^2} \bigr]^2 +  
 f^2 \bigl(  \frac{(z-y) x^2 }{(1-x)^2} \bigr)    }.  
 \end{eqnarray}
The integral equation (\ref{eq:integral}) can easily be solved numerically.
In practice we obtain convergence for
any reasonable initial guess for the function $f(z)$.
It turns out that the ansatz
 \begin{equation}
f ( z )  \approx  \frac{1}{1 + f_2  z^2 }
 \label{eq:ansatz}
 \end{equation}
with $f_2 = {\cal{O}} ( 1 )$, does lead to a rather fast convergence
after a few iterations.
The solution of the integral equation (\ref{eq:integral})
is represented by the solid line in Fig.~\ref{fig:fint}.
In fact, for $f_2 =0.22$
the ansatz (\ref{eq:ansatz})  is already a reasonable
approximation to the solution of Eq.~(\ref{eq:integral})
in the regime $ | z| \lesssim 5$.
Because the quadratic $z$-dependence of the exact solution for small $z$
is correctly described by Eq.~(\ref{eq:ansatz}), this ansatz describes the
spectral function in the vicinity of the quasi-particle peaks quite accurately.
On the other hand, as shown in the inset of Fig.~\ref{fig:fint}, for large
$| z |$ the numerical solution of Eq.~(\ref{eq:integral}) decays as 
$ 1/ | z |$, which is not correctly described by our ansatz ~(\ref{eq:ansatz}). 
The tails of the spectral function are therefore better described by
the interpolation formula 
\begin{equation}
 f ( z ) \approx  \frac{1}{1 +     f_1  | z | },
 \label{eq:fasym}
 \end{equation}
which for $f_1 \approx 0.58$
has the correct asymptotics for large $|z|$, but is less accurate
than Eq.~(\ref{eq:ansatz}) for small $z$.

Given our numerical solution  $f (z)$ of the integral equation (\ref{eq:integral}),
we obtain
 \begin{equation}
 I_0 [ f ] \approx 0.78,
 \end{equation}
and hence
 \begin{equation}
 f_0 \approx \frac{0.66}{m} \sqrt{ \frac{k_0}{ 2 \pi \rho} } .
 \end{equation}
We conclude that for small wavevectors
the on-shell damping of the phase mode
in the one-dimensional Bose gas in units of
its energy $E_k \approx c | k |$  can be written as
 \begin{equation} 
\frac{\gamma_k}{ E_k } \approx 1.32 
\sqrt{ \frac{k_0}{ 2 \pi \rho} }  \frac{ | k |}{ k_0 }.
 \label{eq:gamma1}
 \end{equation}
Note that the dimensionless ratio 
$k_0 / 2 \pi \rho = mc / \pi \rho = u_0 / \pi c$ can be identified
with the Lieb-Liniger parameter divided by $\pi$.
Keeping in mind that in the derivation of 
Eq.~(\ref{eq:gamma1}) we have neglected vertex corrections,
we expect that the prefactor in Eq.~(\ref{eq:gamma1})
is accurate as long as the Lieb-Liniger parameter $u_0/c$ is small.
Comparing  Eq.~(\ref{eq:gamma1})  with the corresponding
expression (\ref{eq:Beliaev}) in $D > 1$
we see that in one dimension
the damping involves an additional factor of $|k | / k_0$; however,
in $D=1$ the
prefactor is proportional to the square root
of the Lieb-Liniger parameter $u_0 / c$, whereas in $D> 1$ it is linear
in the corresponding dimensionless parameter $k_0^D / 2 \pi \rho$.

\begin{figure}[tb]
  \centering
\includegraphics[width=0.45\textwidth]{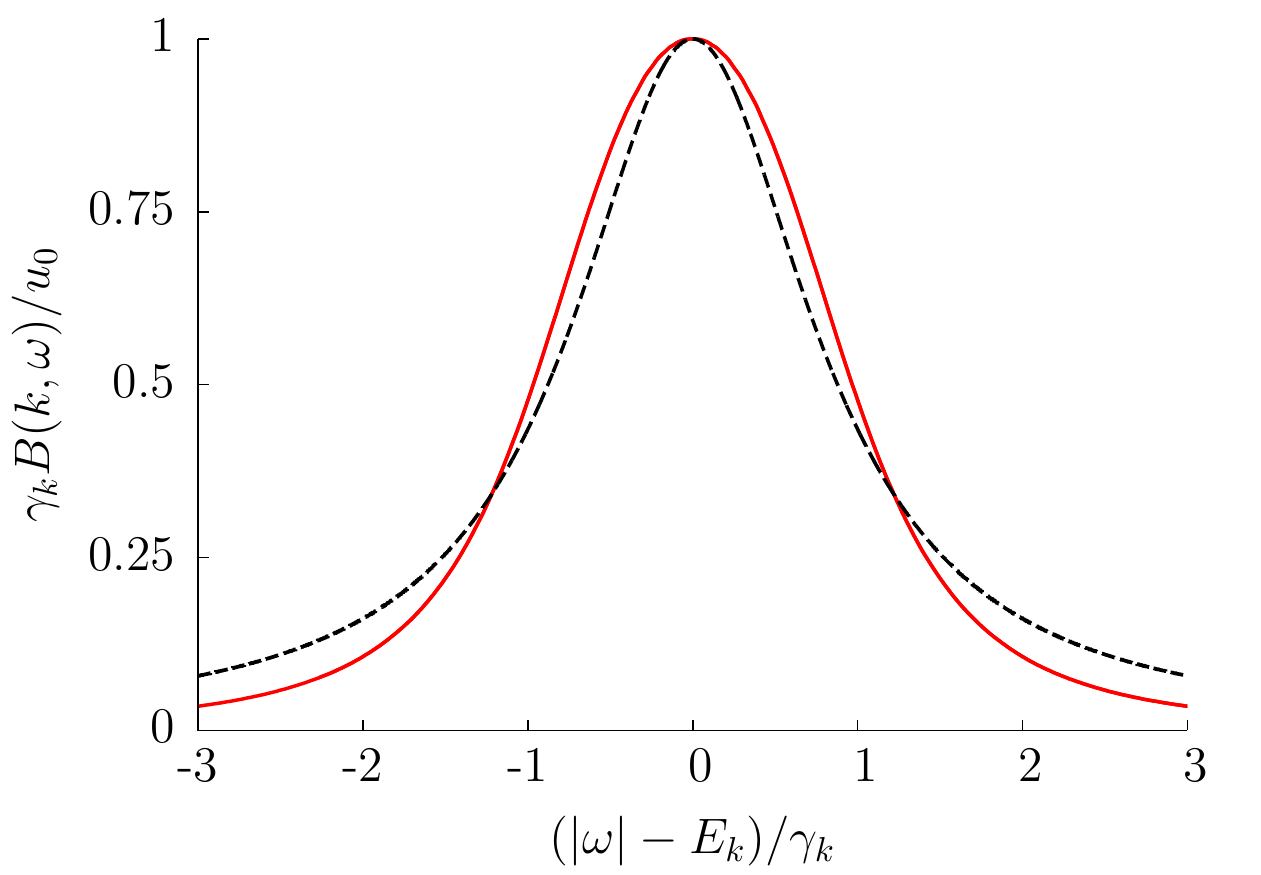}
  \caption{
Normalized spectral function $\gamma_k B ( k , \omega ) / u_0$ of phase fluctuations
of the one-dimensional Bose gas as a function of frequency
for fixed $k=0.5 \, k_{0}$ and 
Lieb-Liniger parameter $u_0/c =0.1$. The on-shell damping is in this case
$\gamma_k/ ( c k_0 ) = 0.059$. 
The dashed line represents a fit to a Lorentzian with on-shell damping $\gamma_k$.
}
\label{fig:Aplot}
\end{figure}

Since the solution of the integral equation
(\ref{eq:integral}) gives the entire scaling function $ f ( z )$
in Eq.~(\ref{eq:gammascaling}), it is now easy to obtain
the momentum- and frequency dependent 
spectral function $B ( k , \omega )$ of phase fluctuations of the one-dimensional
Bose gas. The result is plotted in Fig.~\ref{fig:Aplot}.
Obviously, for frequencies not too far away from the central peak 
($| | \omega | - E_k  | \lesssim 2 \gamma_{k}$)  the line-shape
can be approximated by a Lorentzian, but outside this regime
the spectral function decays faster.
Using the fact that $f ( z ) \sim  ( f_1 | z |)^{-1}  $ for large $z$ we find
that the tails of the spectral function are
 \begin{equation}
 B ( k , \omega ) \sim  \frac{  u_0 \gamma ( k , \omega )}{ (| \omega | - E_k )^2}
 \sim  \frac{1}{f_1}  \frac{  u_0 \gamma_k^2}{   | | \omega | - E_k |^3},
\label{eq:spectralasymptotics}
 \end{equation}
which decays faster than a Lorentzian by a factor of $ \gamma_k / | | \omega |
 - E_k |$. Note that our ansatz (\ref{eq:gammascaling}) is only justified for $\left| \left| \omega \right| - E_{k}\right| \lesssim E_{k}$
so that the result (\ref{eq:spectralasymptotics}) does not describe the asymptotics for $\left|\omega\right|\rightarrow \infty$.

\section{Summary and Conclusions}
\label{Sum}

In summary, we have derived an effective action describing the
dynamics of low-energy and long-wavelength 
phase fluctuations of superfluid  bosons.
Using this action, we have then calculated the
leading momentum dependence of the damping
of the phase fluctuations in arbitrary dimensions.
For $D > 1$  a simple perturbative calculation yields
the usual Beliaev damping,
which scales as $| {\bd{k}} |^{2D-1}$ in $D$ dimensions.
For $ D \rightarrow 1$ the prefactor of $|{ \bd{k}} |^{2D-1}$ vanishes,
and the damping is proportional to $k^2$.
We have obtained this result by taking the 
damping of the intermediate states in the loop integration self-consistently 
into account. In one dimension, we have also calculated the spectral function
of phase fluctuations, which has a Lorentzian line-shape
for frequencies close to the quasi-particle peaks 
associated with the sound mode, but for larger deviations from the
peaks decays faster than a Lorentzian.

Since the vertices of the effective action for the phase fluctuations
vanish for zero wavevectors or frequencies, we believe
that higher orders in perturbation theory do not qualitatively
modify our results. In particular, in one dimension the
spectral function of phase fluctuations does not contain any
algebraic singularity, in contrast
to the spectral function of the amplitude fluctuations \cite{Khodas07,Pustilnik06}.
We are not aware of  experimental methods to directly measure 
the spectral function of phase fluctuations,
so that we cannot compare our result for the
spectral line-shape with experiments.
However, for non-perturbative  calculations of 
the single-particle Greens function of superfluid bosons 
one usually assumes that the Gaussian approximation
is sufficient to calculate 
the propagator of the phase fluctuations  \cite{Popov83,Khodas07}.
Our results imply that
the Gaussian  approximation is indeed well justified in this case, 
because the damping of the phase mode
is small, so that in the superfluid state the phase fluctuations
can propagate as well-defined quasi-particles, even in one dimension.

\section*{ACKNOWLEDGMENTS}
We thank Aldo Isidori and Andr\'{e}  K\"{o}mpel for useful discussions.
This work was financially supported by the DFG via SFB/TRR 49.

\end{document}